\documentclass[prl,10pt, twocolumn]{revtex4}
\usepackage{amsmath}
\usepackage{graphicx}
\usepackage{afterpage}
\usepackage{times}

\begin{document}


\title{Raman study of the phonon symmetries in BiFeO$_3$ single crystals}

\author{C. Beekman$^1$, A.A. Reijnders$^1$, Y.S. Oh$^2$, S.W. Cheong$^2$ and K.S. Burch$^1$}
\affiliation{$^1$Department
of Physics \& Institute of Optical Sciences, University of
Toronto, 60 St. George Street, Toronto, ON M5S 1A7\\
$^2$Rutgers Center for Emergent Materials and Department of
Physics and Astronomy, Rutgers University, 136 Frelinghuysen Road,
Piscataway, NJ 08854, USA.\\}\email{beekmanc@ornl.gov}

\begin{abstract}
\noindent In Bismuth ferrite (BiFeO$_3$), antiferromagnetic and
ferroelectric order coexist at room temperature, making it of
particular interest for studying magneto-electric coupling. The
mutual control of magnetic and electric properties is very useful
for a wide variety of applications. This has led to an enormous
amount of research into the properties of BiFeO$_3$. Nonetheless,
one of the most fundamental aspects of this material, namely the
symmetries of the lattice vibrations, remains controversial. We
present a comprehensive Raman study of BiFeO$_3$
single crystals with the novel approach of monitoring the Raman
spectra while rotating the polarization direction of the
excitation laser. Our method results in unambiguous assignment of
the phonon symmetries, and explains the origin of the controversy
in the literature. Furthermore, it provides access to the Raman
tensor elements enabling direct comparison with theoretical
calculations. Hence, this allows the study of symmetry breaking
and coupling mechanisms in a wide range of complex materials and
may lead to a non-invasive, all-optical method to determine the
orientation and magnitude of the ferroelectric polarization.
\end{abstract}

 \maketitle \noindent

Multiferroic BiFeO$_3$ (BFO) is one of the few materials that
simultaneously exhibits a robust magnetic ordering and large
spontaneous ferroelectric polarization at room
temperature\cite{ramesh}, making it of particular interest for
studying magneto-electric coupling \cite{kimura, hill, caza}. The
mutual control of magnetic and electric properties is of great
interest for applications in spintronics and magnetic storage
media \cite{mathur}. This has triggered significant interest in
BFO, resulting in numerous studies including  optical \cite{lobo,
musfeldt}, and Raman spectroscopy\cite{singh, palai1,porporati},
theoretical calculations\cite{hermet, neaton}, thin film devices
\cite{schlom, ramesh} and electrical control of magnetic
excitations\cite{lee, lebeugle, ren,cazayous1,lobo1}. Amongst
these various techniques, the Raman spectrum of BFO is one of the
most widely studied as it is a powerful tool to investigate
phonons, magnons and their interaction (i.e.
electromagnons).\cite{luke,ren,cazayous1,lobo1} Moreover, proper
phonon mode assignment is necessary to describe the phonons
critical for the multiferroic behavior. However, even for
measurements taken along the high symmetry directions of single
crystals, controversy in the symmetry assignments of the phonon
modes remains. The discrepancies have previously been ascribed to
violation of Raman selection rules due to variations in strain
fields\cite{palai1} (i.e multidomain states) caused by polishing
of the crystal surface. Once the symmetries are unambiguously
assigned, deviations in phonon mode behaviors could be used to
detect the presence of symmetry breaking, multidomain states and
phonon-magnon interactions. Furthermore, simply determining the
mode symmetry only allows for a qualitative comparison with
theoretical calculations. Whereas a quantitative comparison is
enabled by measuring the Raman tensor elements.\\
\indent To this end we have performed a comprehensive set of
polarized micro-Raman spectroscopic studies of BFO single crystals
with uniform ferroelectric polarization. Careful examination and
proper modeling of the rotational dependence of the Raman
intensity enables us to unambiguously assign the (A$_1$, E$_x$ and
E$_y$) modes. Furthermore, we use the presented model to show that
slight misalignment of the crystal leads to ambiguity in the
symmetry assignments. Indeed, our data reveal that comparison of
spectra obtained for different scattering geometries at a single
polarization vector of the incoming light is not sufficient to
have truly unambiguous mode assignment. Nonetheless, unambiguous
assignment can be reached on the as grown single crystal when the
Raman mode intensities as function of crystal rotation are
measured (consistent with previous work on
sapphire\cite{pezzotti}). Hence, with the presented method
polishing is omitted and the resulting ambiguity from misalignment can be avoided.\\
\indent The as-grown BFO single crystals used in this work have
pseudocubic [100]$_{pc}$ facets with a ferroelectric single domain
state\cite{choi} (see Supplemental Material\cite{supp}).   The crystal structure of
BFO (rhombohedral distorted perovskite, R3c) shows a transition
from high to low symmetry accompanied by the formation of
spontaneous electric polarization below the transition temperature
T$_C \sim$1100 K\cite{scott}.  The ferroelectricity is ascribed to
lattice distortions (i.e. off-centering of the Bi-ions) and
results from softening and subsequent freezing of the lowest
frequency polar-phonon mode. The antiferromagnetic ordering sets
in below T$_N\sim$ 640 K with a large magnetic moment of 4 $\mu_B$
on the Fe-ions. Canting of the spins leads to a cycloidal spin
structure with large period (62 nm)\cite{scott,ram} rotating in
the plane containing the electric polarization vector $\textbf{P}$
and cycloid wavevector $\textit{q}$. At room temperature BFO has a
perovskite pseudocubic unit cell (a $\sim$ 3.96 $\dot{A}$)
elongated along the (111)$_{pc}$ direction coinciding with
\textbf{P}. The point group is C$_{3v}$, with 13 Raman active
modes, of which four have A$_1$ symmetry (i.e propagate along the
c-axis) and nine have either E$_x$ or E$_y$ symmetry (i.e.
propagate in the x-y plane), which are doubly degenerate. When the
laser is not along the c-axis, phonons can propagate in the x-z plane, which could lead to LO-TO splitting (i.e. lifts the
degeneracy)
and hence, the presence of A(TO) modes in the \textit{\textit{XX}} and E(LO) modes in the \textit{XY} geometry\cite{loudon,hlinka,stone}, which further complicates the analysis.

\begin{figure}[!htb]
  \includegraphics[totalheight=0.25\textheight,
width=0.42\textwidth,viewport=1 10 610
450,clip]{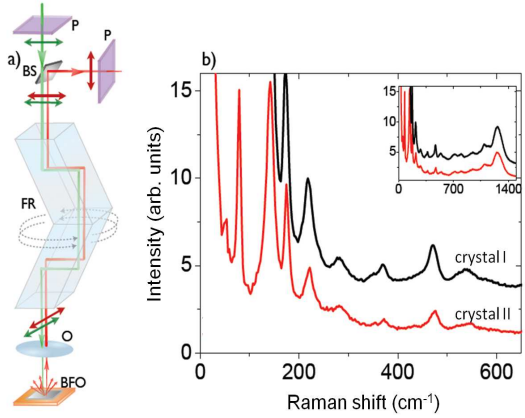}
  \caption{\footnotesize a) Experimental setup with the Fresnel Rhomb (FR) used to rotate the polarization of the
  incoming laser. The green beam is the excitation laser and the red the Raman scattered light, with polarizers (P), Notch filter (BS), objective (O) to focus down the laser and the sample (BFO).   b) Typical single phonon spectra in \textit{XX} geometry for two different [100]$_{pc}$ crystals (black: crystal I and red: crystal II) for Raman shifts between 0 and 650 cm$^{-1}$. Inset: full range up to 1500 cm$^{-1}$. The curves are vertically translated for clarity.}\label{fig1}
\end{figure}

    The Raman spectra were taken
in a backscattering configuration with a Horiba Jobin Yvon LabRam
microscope with a 532 nm excitation source and a 100x objective
(0.8 NA), resulting in a collection area of $\sim$1 $\mu$m (see
Supplemental Material \cite{supp}). All data presented in this work are taken at room
temperature. Furthermore, we investigate the polarization
dependence of the Raman spectra by linearly polarizing the
excitation laser in the plane of the sample and rotating the
polarization direction with steps of 10 degrees over a total of
180 degrees. The rotation is accomplished via a $\lambda$/2
Fresnel Rhomb and is fully equivalent to an in-plane rotation of
the sample (see Fig. \ref{fig1}a and Supplemental Material\cite{supp}). A second
polarizer is used to analyze the scattered light, which is  either
parallel (\textit{XX}) or perpendicular (\textit{XY}) to the incoming polarization
direction.

Fig. \ref{fig1}b shows typical Raman spectra (\textit{XX}  scattering
geometry) taken on two different crystals (both with a
[100]$_{pc}$ surface). The modes below 600 cm$^{-1}$ are single
phonon modes and the broad features above 600 cm$^{-1}$ (see inset
Fig.\ref{fig1}b) are ascribed to 2 phonon excitations,
which is  in agreement with previous reports \cite{palai1}. The
spectrum taken on crystal I shows a total of 11 single phonon
modes (see Table \ref{table}), while crystal II shows a total of
13 single phonon modes (i.e. all modes observed in crystal I and
two additional modes at 53 and 77 cm$^{-1}$, which can be seen due
to the use of a better filter with a lower cutoff frequency).
Raman intensities taken on different locations on one crystal and
on different crystals (Fig. \ref{fig1}b) show similar polarization
dependencies (i.e the symmetry assignments are consistent), which
confirms the single domain character of the crystals. By comparing
the polarization dependence of the Raman intensities of crystal I
and II we show how a different (but homogeneous) ferroelectric
polarization direction influences the phonon mode behaviors
(discussed in detail below).
\begin{figure}[!htb]
  \includegraphics[angle=90,totalheight=0.35\textheight,
width=0.38\textwidth,viewport=10 10 610
450,clip]{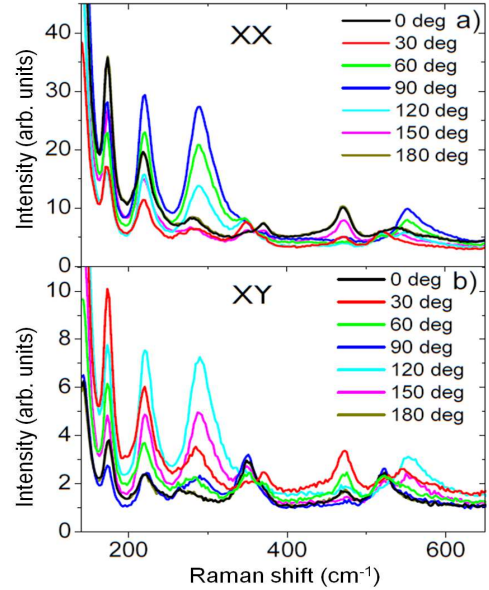}
  \caption{\footnotesize The evolution of Raman spectra as the Fresnel Rhomb is rotated. a) \textit{XX} scattering geometry. b) \textit{XY} scattering geometry. The spectra were taken on the as-grown [100]$_{pc}$ surface of crystal I.}\label{fig2}
\end{figure}
In Fig. \ref{fig2} we show the evolution of the Raman spectra as a
function of in-plane crystal rotation (i.e. rotation of the
polarization direction) taken on the [100]$_{pc}$ surface of
crystal I for the \textit{XX} (Fig.\ref{fig2}a) and \textit{XY} (Fig.\ref{fig2}b)
scattering geometries. Furthermore, we normalize the Raman spectra
at 1500 cm$^{-1}$ to correct for any power fluctuations of the
laser and for polarization dependence of the reflectivity of the
crystal. We have also confirmed that the anisotropy of the optical
constants \cite{choi2} does not significantly influence
polarization dependence of the Raman spectra (see Supplemental Material \cite{supp}).
To quantitatively analyze the data the spectra are fit with
multiple Lorentzian oscillators of the form: $I(\omega) = I_0 +
\sum_i (\frac{A_i\Gamma_{i}}{(4(\omega-E_{i})^2 + \Gamma_{i}^2)})$
where $i$ is the peak number, $I_0$ accounts for the background,
$E_{i}$ is the center frequency, $\Gamma_{i}$ is the width, and
$A_{i}$ is the area of peak $i$. The fitting is done with fixed
mode positions, extracting mode peak intensity (I$_i$(E$_i$)) from
the ratio between area and width of the fitted oscillators (i.e.
I$_i$(E$_i$)=A$_i$/$\Gamma_i$).
\begin{figure}[htb!]
 \includegraphics[angle=90,totalheight=0.45\textheight,
width=0.55\textwidth,viewport=-50 -100 500
430,clip]{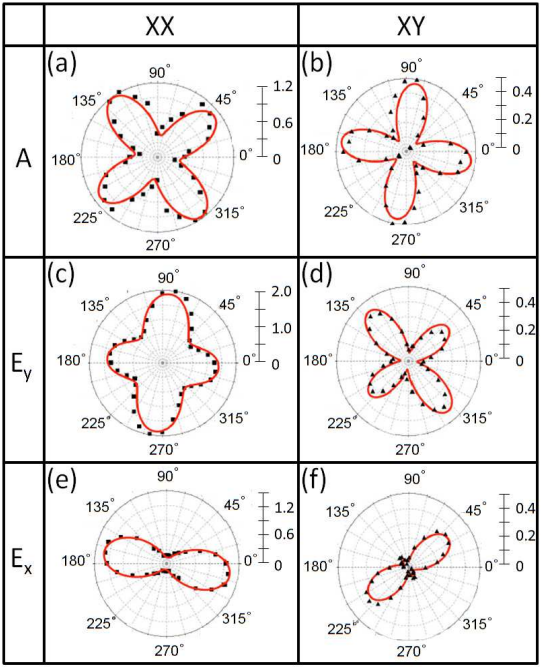}\vspace{-1cm}
\centering\caption{\footnotesize Polar plots of the mode
intensities determined from the Raman spectra (left: \textit{XX} and right:
\textit{XY}) as function of polarization rotation for three representative
modes (a) and b): mode @ 350 cm$^{-1}$, c) and d): mode @ 140
cm$^{-1}$ and e) and f): mode @ 471 cm$^{-1}$) measured on crystal
I. The solid lines are fits ( 350 cm$^{-1}$: A,  140
cm$^{-1}$:E$_y$ and  471 cm$^{-1}$: E$_x$) of which the tensor
elements are indicated in Table \ref{table}. }\label{fig3}
\end{figure}
\begin{figure}[htb!]
 \includegraphics[totalheight=0.40\textheight,
width=0.50\textwidth,viewport=10 -40 660
660,clip]{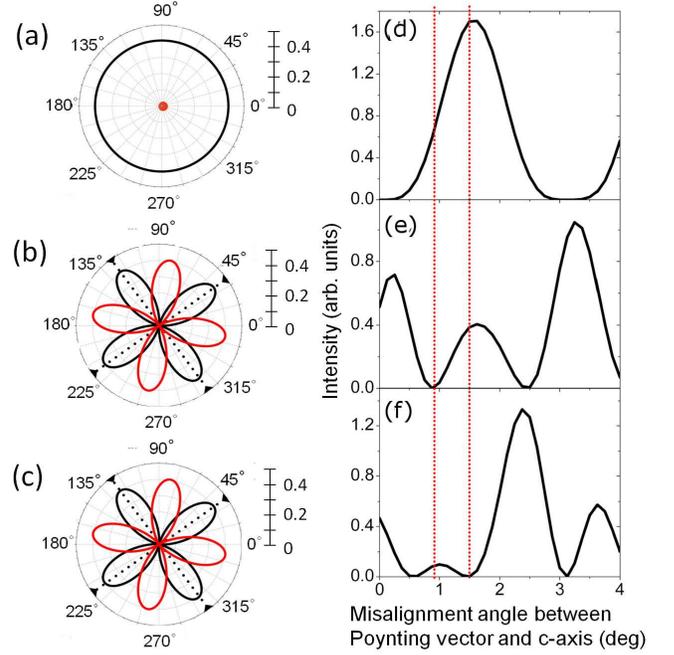}\vspace{-1cm}
\centering\caption{\footnotesize a) - c) Polar plots of calculated
mode intensity variations as function of polarization angle for A
, E$_y$ and E$_x$ symmetry, respectively, in the case of the
perfect alignment of the poynting vector with c-axis (black: \textit{XX},
red: \textit{XY}). The dotted lines in b and c indicate angles for which
E-modes will be mistakingly assigned as A-modes. (d) - f) Linear
plots of the calculated intensity variations in the A, E$_x$ and
E$_y$ modes, respectively (for \textit{XY} scattering geometry) as function
of the deviation of the c-axis from the surface normal. The dotted
lines indicate misalignment angles (0.9 and 1.5$^{\circ}$) for
which the symmetry of 2 out of 3 modes will be misassigned.
}\label{fig4}
\end{figure}
In Fig. \ref{fig3} we show the mode intensities as determined from
fitting the Raman spectra as function of polarization angle for
three representative modes. The polar plots indicate the presence
of  exactly three different mode symmetries. Not surprisingly we
have found that all modes can be sorted into one of these three
types. Indeed, the fits in Fig. \ref{fig3} show that these three
types match well with what we expect for the A, E$_x$ and E$_y$
symmetries. Here we note that  the differences between the mode
behaviors  can be subtle, for example the \textit{XY} curves for the A and
the E$_y$ modes (see Fig. \ref{fig3} b and d) look similar. Hence,
simultaneous modelling of the full polarization curves for both \textit{XX}
and \textit{XY} is necessary and only then results in unambiguous
assignment of the modes.

Moreover, we find no evidence that we are probing phonons that
propagate in the x-z plane (i.e. oblique
phonons\cite{hlinka,stone}). Indeed, such modes would exhibit
LO-TO splitting as seen previously, with the presence of A(TO) and
E(LO) modes leading to reduced intensities of the A(LO) and E(TO)
modes\cite{stone}. Hence, we observe the 13 modes expected from
group theory and the modeling shows that the phonons transform
according to the zone center modes irreducible representations.
Furthermore, on a polished c-axis surface we find the same number
of modes as for the as grown surface with the modes at the same
frequencies (within our resolution).

With a closer look at the model we used for the fits in Fig.
\ref{fig3} we can explain why there is controversy in the phonon
mode assignment in the literature. The extracted mode intensities
as function of polarization angle are modelled using the Raman
tensors for the C$_{3v}$ point group (i.e. the trigonal symmetry
of the lattice). The Raman intensity as function of polarization
angle can be calculated using the equation\cite{pezzotti,Moreira},
\begin{equation} \label{eq1}
I = |e_s^{\dag}R^{\dag}\alpha R e_i|^2
\end{equation}
in which $\alpha$ are the C$_{3v}$ Raman tensors in a trigonal
basis\cite{Hayes},
\begin{align}
A &= \begin{pmatrix}
   a &0 & 0\\
   0 & a & 0 \\
   0 & 0 & b
\end{pmatrix}
&
E_x&= \begin{pmatrix}
  0 &d & 0 \\
d & 0 & e \\
0 & f & 0
\end{pmatrix}
&
E_y&= \begin{pmatrix}
d &0 & -e\\
0 & -d & 0 \\
-f& 0 & 0
\end{pmatrix}
\end{align}
with $R$ the matrix that rotates from cubic to the trigonal orientation,
and with $e_s$ and e$_i$ the polarization vectors that describe
the scattered and incoming light, respectively. The polar plots
for the calculated Raman mode intensity as a function of
polarization rotation are shown in Figs. \ref{fig4}a-c. Here we
have assumed the poynting vector is perfectly parrallel with the
c-axis [111] of the crystal resulting in easy to distinguish
behaviors of the modes. Indeed, in this case determination of A
modes should be easy (Fig. \ref{fig4}a); their intensity is
independent of the polarization angle and should have no
measurable intensity in the \textit{XY} geometry. However, at certain
angles of crystal orientation (indicated by dotted lines in Figs.
\ref{fig4}b and c one can still mistake an E mode for an A mode
(i.e. the mode has intensity in the \textit{XX} geometry but disappears in
the \textit{XY} geometry).  Additional error can come from slight
misalignment of the c-axis with respect to the propagation of the
Raman laser light (i.e. surface normal). Figs. \ref{fig4}d-f
demonstrate that, for the A, E$_x$ and E$_y$ modes respectively,
introduction of a few degrees misalignment can lead to large
variations in the intensity of the phonon modes (in the \textit{XY}
geometry), and hence crossover in mode symmetry assignments. The
two dotted lines in Figs. \ref{fig4}d-f  are examples of
misalignment angles (0.9 and 1.5$^{\circ}$ respectively) for which
the symmetry of 2 out of 3 modes will be misassigned. Also, the
Raman beam is typically focused down, resulting in an average of
incident angles, which already introduces some misalignment.
Here we note that our method results in the same mode symmetry
assignments compared to Palai et al.\cite{palai1}, measured on a
polished c-axis surface. However, we also show that the standard
method of just monitoring the disappearance of modes when
switching from \textit{XX} to \textit{XY} scattering geometry does not provide
adequate information to unambiguously assign the phonon mode
symmetries. Our results not only unambiguously determine the mode
symmetries but also explains the controversy in the literature and
leads to direct determination of the Raman tensor elements.


In Table \ref{table} we show the observed phonon mode frequencies,
Raman tensor elements (\textit{a}, \textit{b}, \textit{d},
\textit{e} and \textit{f}) as obtained from the fits and and their
symmetry assignments for the [100]$_{pc}$ surface of crystal I. We
provide the corresponding polarization curves in the Supplemental
Material\cite{supp}. Moreover, on crystal II, 13 modes were observed (i.e.
the correct amount according to group theory) at the same
frequencies and with the same assignments as presented in Table
\ref{table}; two additional modes were observed at 53 and 77
cm$^{-1}$, the tensor elements and symmetry assignments for all
modes observed on crystal II are shown in Table TI in the
Supplemental Material\cite{supp}. Here we note that the symmetry assignment
of the mode at 279 cm$^{-1}$ remains challenging, because it is
very weak and shouldering the very strong E$_y$ mode at 288
cm$^{-1}$.We have also checked the Raman spectra on a polished
[111]$_{pc}$ (i.e. c-axis) surface (data not shown) on which we
observed a total of 13 modes (the mode at 53 cm$^{-1}$ disappeared
while an additional mode appeared at 70 cm$^{-1}$). The mode at 70
cm$^{-1}$ remains unassigned, it probably also exists on the
[100]$_{pc}$ surface but is too weak and close to a strong E-mode
to be clearly visible. Furthermore, it is possible that the mode
at 53 cm$^{-1}$ is indicative of a violation of Raman selection
rules due to symmetry breaking. Modes at this Raman shift have
been previously assigned as A(TO) modes\cite{palai1}, however they
should not be visible in our scattering geometry and we do not see
evidence of the other A(TO) modes in our spectra. Alternatively
this mode may be an electromagnon\cite{cazayous1}. Future low
temperature studies, where the linewidths are narrow, would help
to better assign these modes. Nonetheless, using the presented
method we have unambigously assigned the phonon modes and
extracted the Raman tensor elements providing quantitative
information for direct comparison with theoretical predications.
Furthermore, the ratio between the Raman tensor elements
\textit{a} and \textit{b} are identical for the A-modes observed
on both crystals. However, we do observe some differences between
the ratios of the E-mode tensor elements between the measurements
taken on crystal I and II, which in no way influences the
consistency of the symmetry assignments. These differences may
indicate that the two crystals (both are single domain) have a
different direction and/or magnitude of the ferroelectric
polarization. This would indeed affect the E-modes but not the
A-modes, since the A-modes are fully symmetrical and constitute
vibrations along the c-axis (i.e. parallel to the ferroelectric
polarization direction). However, this could mean that changes in
the ferroelectric polarization direction leave the mode symmetries
unaltered. Hence, one needs the method presented here to observe
this subtle effect (i.e. changes in the tensor element ratios of
the E-modes) of different ferroelectric polarization on the Raman
intensities.
\begin{table}[!htb]
\begin{center} \includegraphics[totalheight=0.25\textheight,
width=0.53\textwidth,viewport=0 1 580
380,clip]{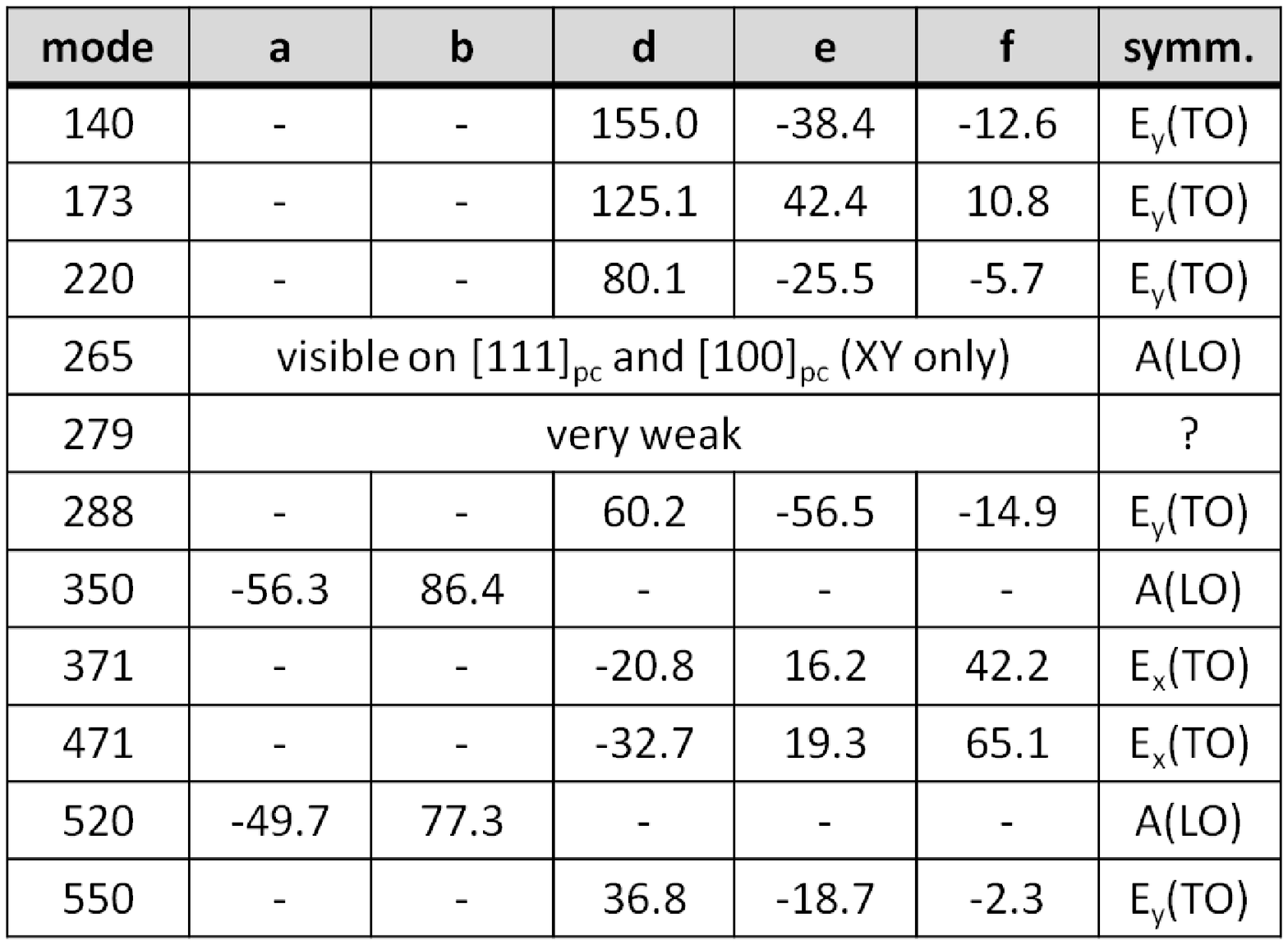}\end{center}
 \caption{\footnotesize Phonon mode frequencies, the Raman tensor elements (\textit{a}, \textit{b}, \textit{d}, \textit{e} and \textit{f}) for the modes as obtained from the fits and the symmetry assignments  for crystal I. The data for all the modes are presented in the Supplemental Material \cite{supp}.}\label{table}
\end{table}\vspace{-0.2cm}


We have measured the evolution of polarized Raman spectra of BFO
single crystals and extracted the polarization curves for every
single phonon mode for both the \textit{XX} and the \textit{XY} scattering geometry.
We fit the \textit{XX} and \textit{XY} curves simultaneously for each mode using a
model based on the Raman tensors of the C$_{3v}$ point group (eq.
\ref{eq1}). As a result unambiguous symmetry assignment and
determination of Raman tensor elements of the phonon modes is
accomplished even on the as grown [1 0 0]$_{pc}$ surface. In Fig.
\ref{fig3} the excellent and unambiguous agreement between the
experimental results and our calculations for a [100]$_{pc}$
surface demonstrate the importance of performing the Raman
measurements over a full rotation of the crystal. Whereas the
calculations in Fig. \ref{fig4} demonstrate that only measuring
the \textit{XX} and \textit{XY} spectra on a [111]$_{pc}$ surface for a single
polarization direction (as is typically done), can easily lead to
wrong assignments of the phonons due to misalignment of the
crystal. It is clear that unambiguous mode assignment can only be
reached if one monitors the Raman signal as function of rotation
of the crystal. Simply comparing \textit{XX} and \textit{XY} scattering geometries
for one polarization angle is not enough even for a c-axis
surface. Besides obtaining unambiguous mode assignment for BFO,
this work has wider implications as well. The method can be used
on any material to check crystal symmetry and assign the phonon
modes. Furthermore, once unambiguous assignment has been
accomplished one can use the presented method to investigate
symmetry breaking as well, for example by studying deviations in
the tensor element ratios, the symmetry assignments and through
observation of more than the predicted number of modes. This gives
us a powerful tool to investigate occurrence of (electro)magnons
and compare to existing reports\cite{cazayous1,lobo1}. Moreover it
would allow study of coupling mechanisms in complex materials such
as multiferroics, as well as provides quantitative information for
direct comparison with theoretical predications. 

We are grateful for numerous discussions with R. de Sousa, N.B Perkins, and H.Y. Kee and we thank Harim Kim for designing Fig. \ref{fig1}a. Work at the University of Toronto was supported by NSERC, CFI, and ORF; work at Rutgers University was supported by the NSF under grant NSF-DMR-1104484.

\normalsize



\vspace{-0.2cm}
\begin{thebibliography}{60}\footnotesize
\bibitem{ramesh} R. Ramesh and N. A. Spaldin, Nat. Mater. \textbf{6}, 21 (2007)
\bibitem{kimura} T. Kimura et al., Nature, \textbf{426}, 55 (2003)
\bibitem{hill} N. A.  Hill, J. Phys. Chem. B \textbf{104}, 6694 (2000)
\bibitem{caza} P. Rovillain, R. de Sousa, Y. Gallais, A. Sacuto, M. A. MŽasson, D. Colson, A. Forget, M. Bibes, A. BarthŽlŽmy, and M. Cazayous, Nature Mater. \textbf{9}, 975 (2010)
\bibitem{mathur} W. Eerenstein, N. D. Mathur and J. F. Scott, Nature \textbf{442}, 759 (2006)
\bibitem{lobo} R. P. S. M. Lobo, R. L. Moreira, D. Lebeugle, and D. Colson, Phys. Rev.  B \textbf{76}, 172105 (2007)
\bibitem{musfeldt} X. S. Xu, T. V. Brinzari, S. Lee, Y. H. Chu, L. W. Martin, A. Kumar, S. McGill, R. C. Rai, R. Ramesh, V. Gopalan, S. W. Cheong, and J. L. Musfeldt, Phys. Rev. B \textbf{79}, 134425 (2009)
\bibitem{singh} M. K. Singh, H. M. Jang, S. Ryu, and M.-H. Jo, Appl. Phys. Lett. \textbf{88}, 042907 (2006)
\bibitem{palai1} R. Palai, H. Schmid, J. F. Scott, and R. S. Katiyar Phys. Rev. B. \textbf{81}, 064110 (2010) and references therein
\bibitem{porporati} A. A. Porporati, K. Tsuji, M. Valant, A-K. Axelssond and G. Pezzotti, J. Raman Spectrosc. \textbf{41}, 84 (2010)
\bibitem{hermet} P. Hermet, M. Goffinet, J. Kreisel, and Ph. Ghosez, Phys. Rev. B \textbf{75}, 220102(R) (2007)
\bibitem{neaton} J. B. Neaton, C. Ederer, U. V. Waghmare, N. A. Spaldin, and K. M. Rabe, Phys. Rev. B \textbf{71}, 014113 (2005)
\bibitem{schlom}  J. Wang, J. B. Neaton,  H. Zheng,  V. Nagarajan, S. B. Ogale, B. Liu, D. Viehland, V. Vaithyanathan,  D. G. Schlom, U. V. Waghmare, N. A. Spaldin, K. M. Rabe, M. Wuttig, R. Ramesh, Science \textbf{299}, 1719 (2003)
\bibitem{lee} S. Lee, W. Ratcliff, S.-W. Cheong, and V. Kiryukhin, Appl. Phys. Lett. \textbf{92}, 192906 (2008)
\bibitem{lebeugle}D. Lebeugle, D. Colson, A. Forget, M. Viret, A. M. Bataille, and A. Goukasov, Phys. Rev. Lett. \textbf{100}, 227602 (2008)
\bibitem{ren}F. Wang, J.-M. Liu, and Z.F. Ren, Adv. Phys. \textbf{58}, 321 (2009)
\bibitem{cazayous1} M. Cazayous, Y. Gallais, and A. Sacuto, R. de Sousa, D. Lebeugle, and D. Colson, Phys. Rev. Lett. \textbf{101}, 037601 (2008)
\bibitem{lobo1} P. Rovillain, M. Cazayous, Y. Gallais, and A. Sacuto, R. P. S. M. Lobo, D. Lebeugle, and D. Colson, Phys. Rev. B \textbf{79}, 180411(R) (2009)
\bibitem{luke} L. J. Sandilands, J. X. Shen, G. M. Chugunov, S. Y. F. Zhao, Shimpei Ono, Yoichi Ando, K. S. Burch, Phys. Rev. B \textbf{82}, 064503 (2010)
\bibitem{pezzotti} M.C. Munisso, W. Zhu, and G. Pezzotti, Phys.
Status Solidi B \textbf{246}, 1893 (2009)
\bibitem{choi} T. Choi, S. Lee, Y. J. Choi, V. Kiryukhin and  S.-W. Cheong, Science \textbf{324}, 63 (2009); S. Lee, T. Choi, W. Ratcliff, R. Erwin, S-W. Cheong, and V. Kiryukhin, Phys. Rev. B \textbf{78}, 100101(R) (2008)
\bibitem{supp}See Supplemental Material for additional data and the table containing the Raman tensor elements and symmetry assignments
for the phonons observed on crystal II.
\bibitem{scott} G. Catalan, and J.F. Scott, Adv. Mater.  \textbf{21}, 2463 (2009)
\bibitem{ram} M. Ramazanoglu, W. Ratcliff, Y. J. Choi, S. Lee, S.-W. Cheong, and V. Kiryukhin, Phys. Rev. B,  \textbf{83}, 174434 (2011) and references therein
\bibitem{loudon} R. Loudon, Adv. Phys. \textbf{13}, 423 (1964)
\bibitem{hlinka} J. Hlinka, J. Pokorny, S. Karimi, and I.M. Reaney,
Phys. Rev. B, \textbf{83}, 020101(R) (2011)
\bibitem{stone} G. Stone, V. Dierolf, Opt. Lett. \textbf{37}, 1032
(2012)
\bibitem{choi2} S. G. Choi, H. T. Yi, S.-W. Cheong, J. N. Hilfiker, R. France, and A. G. Norman, Phys. Rev. B, \textbf{83}, 100101 (2011)
\bibitem{Moreira}R. L. Moreira, M. R. B. Andreeta and A. C. Hernandes and A. Dias, Crystal Growth and Design \textbf{5},  1458 (2005)
\bibitem{Hayes}W. Hayes and R. Loudon, Scattering of light by crystals, John Wiley (1978)
\end{thebibliography}
\end{document}